# How to sustain the terrestrial biosphere in the Anthropocene? A thermodynamic Earth system perspective

Axel Kleidon
Max-Planck-Institut für Biogeochemie
Hans-Knöll-Str. 10
07745 Jena
Germany

Contact e-mail: akleidon@bgc-jena.mpg.de

**Abstract:**
Many aspects of anthropogenic global change, such as land cover change, biodiversity loss and the intensification of agricultural production, threaten the natural biosphere. These aspects seem somewhat disjunct and specific so that it is hard to obtain a bigger picture of what these changes imply and to distinguish beneficial from detrimental human impacts. Here I describe a holistic approach that provides such a bigger picture and use it to understand how the terrestrial biosphere can be sustained in the presence of increased human activities. This approach focuses on the free energy generated by photosynthesis, energy needed to sustain either the dissipative metabolic activity of ecosystems or human activities, with the generation rate being set by the physical constraints of the environment. We can then distinguish two kinds of human impacts on the biosphere: detrimental effects caused by enhanced human consumption of this free energy, and empowering effects that allow for more photosynthetic activity and therefore more dissipative activity of the biosphere. I use examples from the terrestrial biosphere to illustrate this view as well as global datasets to show how this can be estimated. I then discuss how certain aspects of human-made technology can act to enhance the free energy generation of the terrestrial biosphere, which can then facilitate sustaining the biosphere in times at which human activity increasingly shapes the functioning of the Earth system.

# 1. SUSTAINABLE ENERGY AS THE CORE PROBLEM OF THE ANTHROPOCENE

Energy is at the core of many of the global challenges that we currently face, including those that challenge the functioning of the biosphere. What I want to show here is that this focus on energy and how it is converted within the Earth system helps us to get a clearer, big picture of how current human activity inevitably results in a diminished biosphere, but also that with the help of technology, we can make informed choices to better sustain the natural biosphere in the future in which human activities are likely to increasingly shape planetary functioning.

Let us first look at a few examples of how current global challenges relate to energy. An almost obvious example is global warming. The increased consumption of fossil fuels directly relates to the increased need of human societies for energy to fuel their socioeconomic activities. This comes at the inevitable consequence of increased greenhouse gas concentrations in the atmosphere, which causes climate to change. Global warming thus directly relates to human energy consumption.

When we look at tropical deforestation as another example, the link is not quite so obvious. Tropical deforestation is mainly caused by the conversion of natural forests into pastures and cropland. This conversion aims at producing more food, and food relates to the calories that human metabolisms need to be sustained. So an expansion of agricultural areas relate to the increased food production, which is equivalent to energy generation in chemical form suitable to meet human demands for food. So tropical deforestation also directly links to human energy needs.

The loss of biodiversity is the last example I want to use. While there are many factors that are thought of being detrimental to biodiversity (IPBES, 2019), such as land cover change, habitat destruction, and intensified agriculture, at its core is energy as well. Each organism needs chemical energy to sustain its metabolism. This energy comes from photosynthesis, just as it is the case for food production for human societies. There is a suite of hypotheses that explain biodiversity patterns in terms of energy (e.g., see reviews by Currie et al., 2004 and Clarke and Gaston, 2006). Simply speaking, these hypotheses in essence state that tropical regions have more energy available due to their higher productivity, this allows them to sustain the metabolisms of more organisms, and thus higher diversity levels. So when humans convert and use land more intensively for food production, then less energy is left for the metabolic activities of the natural biosphere. Hence, the loss of biodiversity with increased and intensified land use also seems to be a direct consequence of greater human energy demands.

These examples suggest a general dilemma in which human activity increasingly diverts energy from the Earth system to their own use, be it to sustain food demands or socioeconomic activity, with the price being that less is left behind for the natural biosphere to exist (Figure 1). A key component of this dilemma is that the productivity of the biosphere has natural limits set by the environment. This, in turn, sets limits to growth of human societies, as described more detailed in the seminal work by Meadows et al (1972) on the "*Limits to Growth*". When human societies approach this limit, they unavoidably deteriorate the natural biosphere because less is left behind. It seems to imply an unavoidable, tragic outcome of increased energy demands of human societies.

Here, I want to show that we can avoid this tragic outcome. One option seems to be simply to consume less energy, but one may wonder how realistic this option is to accomplish. I want to focus on another option, one that allows for further increases in energy consumption by human societies, but decouples this increase from detrimental impacts this has on the biosphere. This requires us to find ways to enhance photosynthesis beyond its natural limits so that more energy is available for the natural biosphere and human societies, or to identify other technological options to generate the energy needed to sustain socioeconomic activities beyond natural levels (shown as "feedback" in Figure 1).

There are examples for such options from the human past. Irrigation-based agriculture using river water or human-made reservoirs already enhanced agricultural productivity for thousands of years, e.g., along the Nile river in Egypt. By diversion, it makes water available for terrestrial productivity that was not possible by natural means. Looking into the future, more modern types of human-made technology can accomplish this task with much greater impact, particularly by using seawater desalination and photovoltaics. Seawater desalination by membranes is much more efficient in desalination than the natural hydrologic cycle of evaporation and subsequent precipitation, while photovoltaics is much more efficient than natural photosynthesis in generating energy. These technologies can decouple energy and water needs of human societies from the natural system and tremendously boost the availability of energy to human societies. This would then allow for more of the naturally generated energy to be left to sustain the natural biosphere in the future despite growing energy demands of human societies, potentially resulting in a positive feedback (as shown by the dotted line in Figure 1).

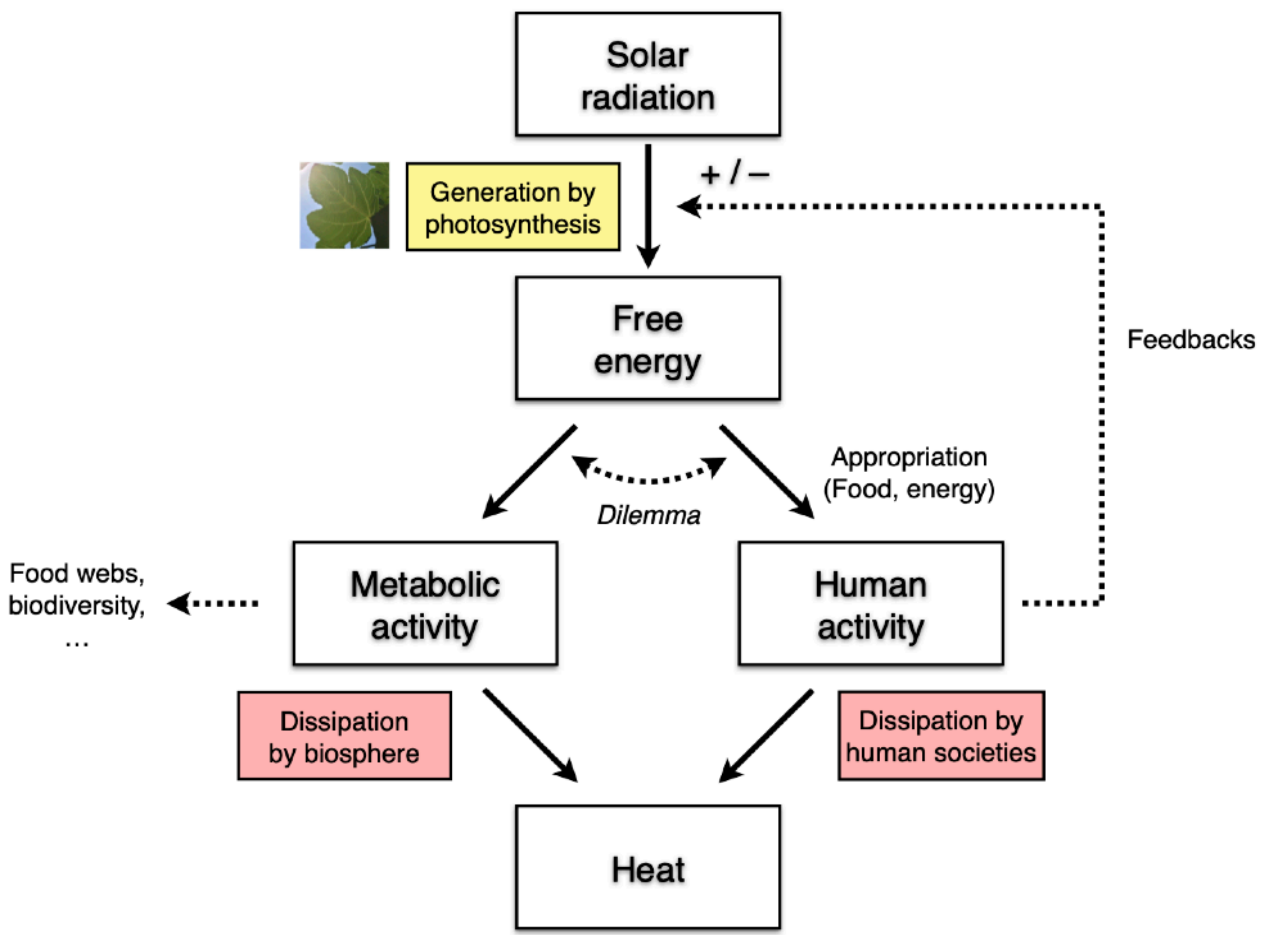

**FIGURE 1:** Schematic diagram to illustrate the basic dilemma related to using the energy generated by photosynthesis to sustain the metabolic activity of the natural biosphere or the activities of human societies. Feedbacks from human activity on photosynthesis can potentially generate more free energy, which could help to avoid the dilemma.

Before I substantiate this more optimistic option for the future, we first need to clarify the use of the term energy. It is important to note that there is an important difference between different forms of energy. The discussion here focuses on the concept of free energy, energy that was generated by work, and that is able to perform further work. Examples for free energy is the kinetic energy in the winds of the atmosphere or river currents, carbohydrates are free energy in chemical form that can fuel metabolic reactions, while power plants and photovoltaics generate free energy in electric form. So the energy that human societies need to feed their metabolisms and fuel their socioeconomic activities is free energy, as is the energy that sustains living organisms. When we want to understand how the activity of the natural biosphere can be sustained in the presence of growing human demands, we need to get into more detail how free energy is being generated from the solar forcing of the planet, and why some human-made technology performs better than photosynthesis or natural desalination.

The following parts of this paper are structured as follows: In the next section (Section 2), I describe how photosynthesis generates free energy from sunlight and provide an explanation why it is has such a low efficiency. This is substantiated with maps that were derived from a simple, physical description of this limitation from previous work and that can be used to provide first-order estimates of the magnitude of free energy generation by the natural, terrestrial biosphere. In Section 3, I then describe an example to illustrate how the biosphere has means to push its physical limits to higher levels, thereby affecting environmental conditions that are more conducive to perform photosynthesis, and thus generating more energy to fuel more activity within the biosphere. This example is used to substantiate the notion that natural systems push their physical limits, which is likely a rather general feature of evolving thermodynamic systems and may apply to human systems as well. In Section 4 I then describe the energy consumption of human societies and provide estimates of how much human activity has already diminished the natural biosphere on land. These estimates substantiate how important human activity has become as an Earth system process in quantitative, physical terms. I then provide a few examples in section 5 on how human-made technology can push limits on human energy consumption which, at the same time, could be used to sustain or enhance the activity of the natural biosphere. I close with a brief summary and conclusions.

## 2. HOW TO GENERATE FREE ENERGY FROM THE SOLAR FORCING

Before we start describing how photosynthesis generates free energy, we need to briefly define this term, describe what makes it so different to "just" energy, and how it is generated by Earth system processes. Free energy is energy without entropy, capable of performing work, which can then result in so-called dissipative dynamics. It is sometimes referred to as exergy (e.g., Hermann, 2006). The kinetic energy associated with atmospheric motion is an example of free energy, which is dissipated by friction, as is the chemical energy stored in carbohydrates and biomass, which is dissipated by metabolisms or combustion. Free energy plays a central role for the dynamics of the Earth system, driving the physical dynamics that shape climate, the biospheric dynamics with its food chains, as well as socioeconomic dynamics. These dynamics are driven by the dissipation (or consumption) of this free energy, forming dissipative systems that are thermodynamically very different to those that are in thermodynamic equilibrium.

To understand how free energy is generated from the solar forcing, we need to look closer at entropy, a key aspect of energy. Entropy was originally introduced empirically at the advent of steam engines in the 19th century to describe how much work can be derived from a heat source. It received a physical interpretation with the work of Boltzmann in the late 19th century and its subsequent extension by Planck to the treatment of radiation, together with the notion that energy at the scale of atoms comes in discrete amounts called quanta. This set the foundation for the revolution of quantum physics in the early 20th century. In modern physics, entropy plays a key role to describe the many facets of the quantum world of atoms and molecules in terms of comparatively simple, macroscopic characteristics that describe how energy is stored and converted in solids, liquids, and gases.

At the microscopic scale of atoms, energy comes in discrete amounts called quanta. Energy in form of radiation comes in form of photons, energy in chemical bonds represent discrete distributions of electrons across different atomic shells, while heat refers to the random motion, vibration, or rotation of molecules. All these microscopic aspects are represented by discrete amounts of energy being distributed over finite number of states. They can thus be counted, and we can assign probabilities to certain ways to distribute these quanta of energy across the states. At the macroscopic scale, however, we are typically not interested in these details. Then, we can make the assumption that a given amount of energy is distributed in the most probable way. This is measured by entropy, as expressed by Boltzmann's famous equation, $S = k \log W$, where $S$ is the entropy, $k$ is a constant, and $W$ is the number of possible ways to distribute energy. The assumption of the most probable distribution represents so-called thermodynamic equilibrium. Since at the microscopic scale energy is distributed across photons, electrons, and molecules, we actually have three forms of entropy that are important to Earth system science: radiative entropy, molar entropy, and thermal entropy.

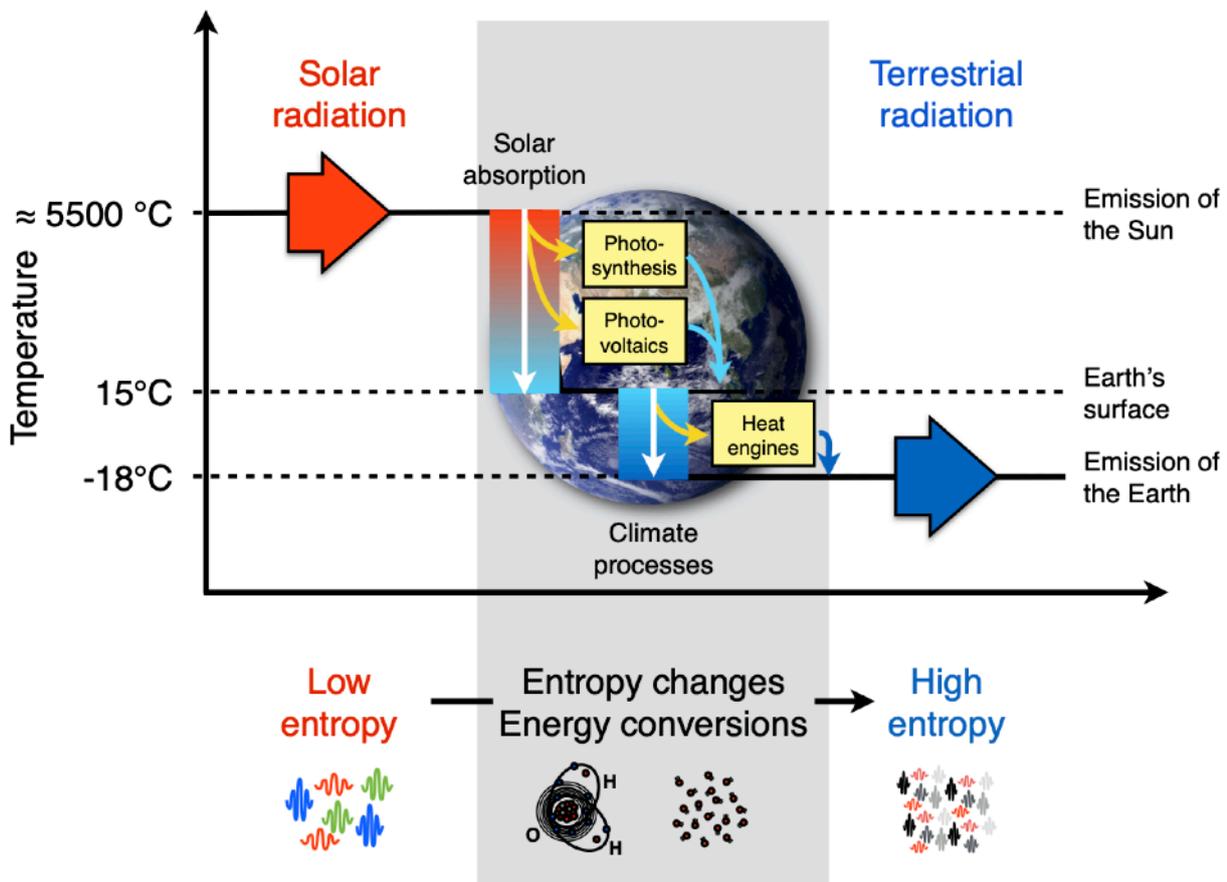

**FIGURE 2:** At the planetary scale, there are three potential ways to derive free energy from low-entropy sunlight: heat engines that drive the dynamics in the physical environment, photosynthesis, and photovoltaics.

Systems become interesting when they are not in equilibrium, and this will bring us to the concept of free energy. For a disequilibrium we need to have differences in entropy. Here, the second law of thermodynamics kicks in, requiring that whatever will happen, it needs to overall yield an increase in entropy. For the Earth system, the major driver for disequilibrium is the difference in the kind of radiation that the Earth receives and emits to space (Figure 2). At the planetary scale, the energy fluxes are roughly balanced, so that about as much solar radiation enters the Earth system as is reflected and emitted to space. But these energy fluxes differ vastly by their radiative entropies. Solar radiation was emitted from the Sun at a very high emission temperature of about 5500 °C, which results in radiation with short wavelengths, mostly in the visible range, and very low radiative entropy at the Earth's orbit. After absorption and further transformations, the Earth emits this energy as terrestrial radiation at a much lower, so-called radiative temperature of about -18°C. This radiation is mostly represented by infrared wavelengths and has a much higher entropy. This results in a massive thermodynamic disequilibrium between the solar radiation the Earth receives and the radiation the Earth emits.

The simplest way to destroy this disequilibrium is to simply absorb and re-emit radiation at a lower temperature. It increases entropy, yet does not drive dissipative dynamics. More relevant are the cases in which this disequilibrium is used to generate free energy. There are different mechanisms of how this can be accomplished, yet the rules for these mechanisms are the same and set by the laws of thermodynamics.

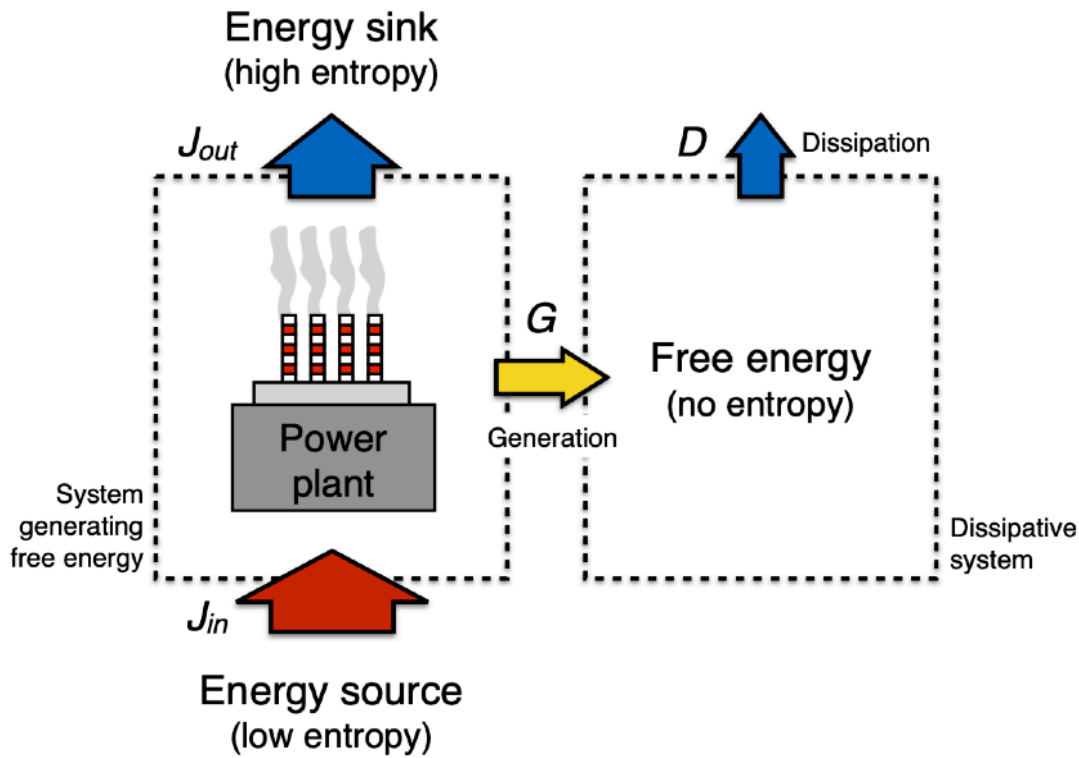

**FIGURE 3:** Illustration of free energy generation using a power plant as an example (left box). The same thermodynamic rules also apply to energy conversions in the Earth system. Once free energy is generated, it drives the dynamics of dissipative systems (box on the right).

The physical way to do this can be illustrated by a conventional power plant (Figure 3). Heat is generated by combustion of a fuel at a high temperature, yielding heat at low entropy. That it has low entropy can be seen by Clausius's expression, which states a change in entropy as $\Delta S = \Delta Q/T$, with $\Delta Q$ being the heat added or removed, and $T$ being the temperature (in units of Kelvin) at which heat is exchanged. Because combustion takes place at a high temperature, the added entropy to the power plant is comparatively small. The steam released by the cooling towers expels some of that heat from the power plant, but at much colder temperatures, thus exporting heat with much higher entropy. To satisfy the second law of thermodynamics, there must be at least as much entropy being released from the cooling towers as is added by combustion. When these entropy fluxes balance each other, with entropy entering the power plant at the same rate as entropy exiting through the cooling towers, this yields the upper limit on how much energy without entropy can be generated, that is, free energy. This limit is very well known as the Carnot limit. It sets the limit to how much work can at best be performed and how much electricity, free energy in electric form, can at best be generated by the power plant.

The physical Earth system operates much like such a power plant. The heat source is the absorption of solar radiation of the surface (instead of being released by combustion), and the emission of radiation from the atmosphere serve as the cooling towers that exports entropy from the Earth system to space in form of radiation. The work done is that of generating motion: either in form of buoyancy, driving vertical convective motions, or horizontally, in form of regional circulations (such as a sea breeze system), and even the large-scale circulations such as the Hadley circulation or the mid-latitude winds. Comparison to observations show that atmospheric motion in fact operates at this thermodynamic limit, working as hard as it can (Kleidon, 2021a). This maximisation of power is reflected in characteristic surface energy balance partitioning, temperature patterns, and evaporation rates that compare very well with observations (Kleidon, 2021b).

Motion then drives other physical processes, such as generating waves over the ocean, hydrologic cycling, renewable wind energy generation, or it is dissipated back into heat by friction. The work involved is, however, relatively small, and the conversion has a low efficiency. This is because only differences in radiative heating serve as the heat source, and the temperature differences are much smaller compared to that of a power plant. This amounts in a low overall conversion efficiency of less than 1% of the incoming solar radiation being converted to free energy in form of kinetic energy.

This low conversion efficiency for physical Earth system processes is inevitable. Once solar radiation is absorbed at the Earth's surface and converted into heat, most of its low entropy is already lost, because the surface is at a much colder temperature than the emission temperature of the Sun. Absorption thus turns solar radiation into heat of relatively high entropy. The temperature differences for converting this energy further are thus set by the difference between the surface and the Earth's radiative temperature, or between the tropics and polar regions. This difference is quite small (about 33 K), yielding the low conversion efficiency.

To make better use of solar radiation, it requires mechanisms that avoid the intermediate conversion step into heat and rather turn solar energy directly into free energy instead of heat. There are two of such alternatives, indicated by the yellow boxes in Figure 2: photosynthesis and photovoltaics. We next turn to photosynthesis, as this is the process by which free energy is generated from sunlight for the dissipative activity of the biosphere.

## 3. ENERGY GENERATION BY THE NATURAL BIOSPHERE AND ITS PHYSICAL LIMITS

To evaluate the biosphere using this thermodynamic view, the key question is how and how much free energy can be generated by photosynthesis, which then constrains the level of metabolic activity within the biosphere. Typically, photosynthesis is described as a chemical conversion process which converts carbon dioxide and water into carbohydrates and oxygen, using solar radiation as the energy source. The resulting carbohydrates then contain about 40 kJ of chemical free energy per gram of carbon. This energy feeds the metabolic activities of the producers, known as autotrophic respiration, as well as of living organisms, or heterotrophic respiration, that make up the biosphere (Figure 4). This metabolic activity uses the chemical free energy contained in the organic carbon compounds generated by photosynthesis as well as oxygen, and dissipates this free energy back into heat, thereby producing entropy. While the focus here on photosynthesis does not tell us how and how many organisms are being fed by this chemical free energy, its generation nevertheless creates thermodynamic disequilibrium - in form of reduced, organic carbon compounds and atmospheric oxygen - and it sets the magnitude for the dissipative activity of the biosphere.

In the following, we first look at the energy conversions that are involved in photosynthesis in somewhat greater detail, estimate their conversion efficiencies, and evaluate whether these operate at their thermodynamic limit, just as atmospheric motion is in the climate system.

The first step of photosynthesis involves the so-called light reactions in the photosystems during which light is absorbed. Here, light does not turn into heat - random motion of molecules - but rather performs photochemistry as it splits water into its compounds, and further splits hydrogen into its negatively charged electron and its positively charged nucleus. In other words, the photosystems perform the work of charge separation, generating electric free energy. Photosynthesis requires about 8 to 10 quanta of light of wavelengths of about 700 nm to split the hydrogen atoms involved in binding one molecule of carbon, described by the well-established concept of quantum yield efficiency (Emerson, 1958). These quanta carry about 1.8 eV of energy each, with 1 eV = 1.6 x $10^{-19}$ J being a unit of energy at the quantum scale and the amount calculated by $h c / \lambda$, with $h \approx 6.63$ x $10^{-34}$ J s being the Planck constant, $c \approx 3 \times 10^8$ m s$^{-1}$ the speed of light, and $\lambda = 700 \times 10^{-9}$ m being the wavelength of the photon. Taken together, this yields energy from absorbed radiation of about 14.4 - 18 eV. For comparison: this amount is slightly more than the bare minimum of 13.6 eV needed to perform the work of charge separation of the hydrogen atom. On a mol basis, photosynthesis uses at least $N_a \times$ 14.4 eV = 1387 kJ mol$^{-1}$ to split one mol of water, with $N_a$ being the Avogadro constant, $N_a = 6.022 \times 10^{23}$ mol$^{-1}$. Hence, this first step is highly efficient, with a conversion efficiency of about 76 - 94 %.

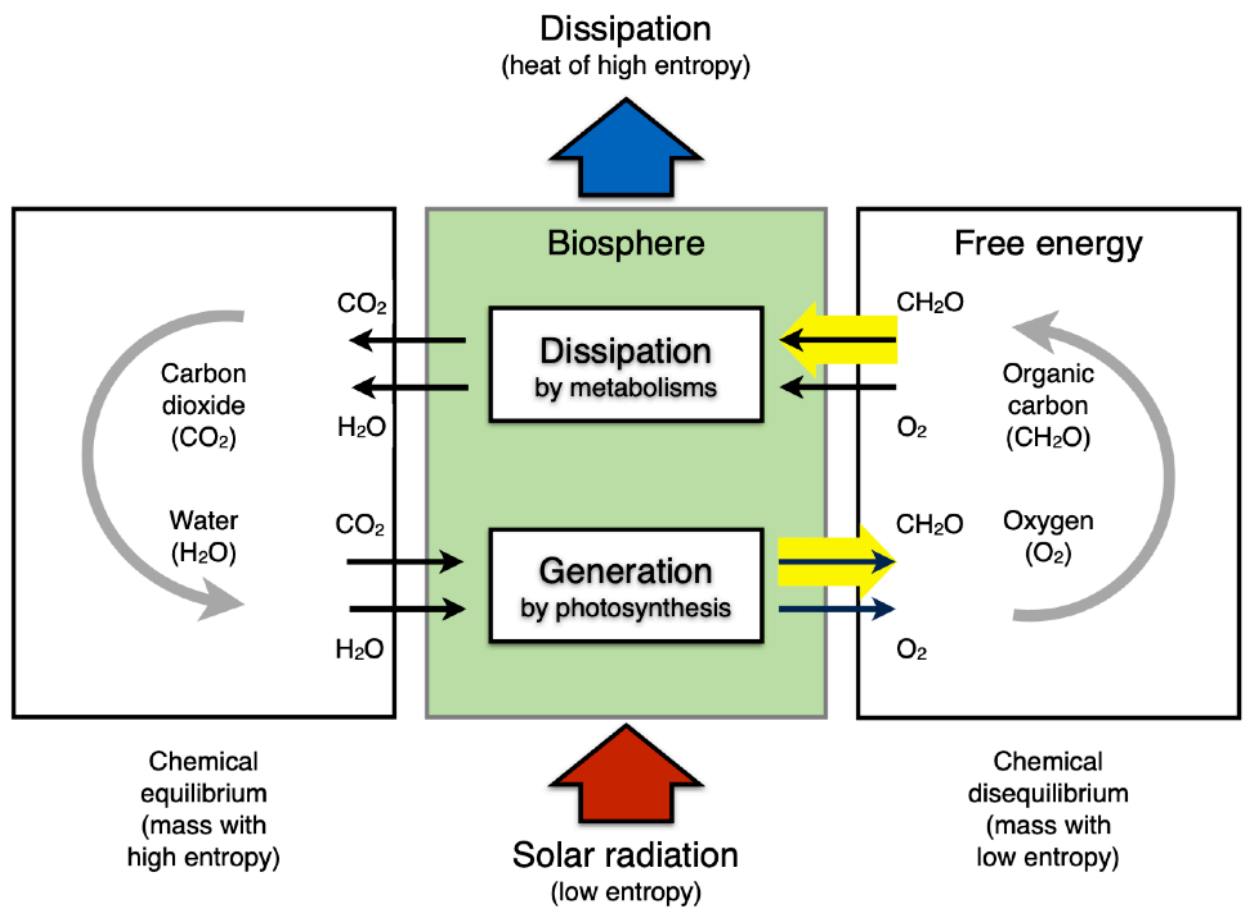

**FIGURE 4:** Schematic diagram of the energetics of the biosphere. The dynamics are driven by the generation of free energy by photosynthesis, which is associated with a chemical disequilibrium of organic, reduced carbon and oxygen. This free energy is dissipated by the metabolic activities of producers and consumers of the biosphere, but also of human societies.

The generated electric energy is then incorporated into longer-lived chemical compounds of NADP and ATP, before these are used in the Calvin cycle to convert this energy further and store it in form of carbohydrates. This requires carbon dioxide, which needs to be taken up from the surrounding air. This step is far less efficient. Using the 1387 kJ of energy of the absorbed photons, this cycle produces one mol of carbon in form of glucose with an energy content of merely 480 kJ. This corresponds to an overall conversion efficiency from radiative to chemical energy of 480 kJ/1387 kJ = 34%. Laboratory measurements at low light conditions found that plants operate close to this efficiency (Hill and Rich, 1983). When we further take into account that photosynthesis can only utilize about 55% of the solar spectrum, the so-called photosynthetically active radiation, or PAR, this reduces the efficiency of carbon fixation to less than 19% for converting the energy contained in sunlight into carbohydrates.

Observations from terrestrial ecosystems, however, show that in general, the efficiency of photosynthetic carbon uptake is substantially lower than this efficiency, with values typically being less than 3% (Monteith, 1972; 1977; Kleidon, 2021b). This much lower efficiency can be attributed to the constraining role of gas exchange associated with carbon and water between the vegetation canopy and the surrounding air (Kleidon, 2021b). Vegetation needs to take up carbon dioxide from the air, and while doing so, it inadvertently loses water vapour. This gas exchange with the atmosphere takes place at a relatively fixed ratio of about 2 grams of carbon taken up for each kg of water evaporated, the so-called water use efficiency (Law et al, 2002). This implies that when we want to identify the primary limitation for photosynthesis, and thus for the free energy generation of the biosphere, we need to understand what limits the gas exchange between the surface and the atmosphere, or, closely associated, the rate of evaporation.

This brings us back to the constraining role of thermodynamics, not in terms of the energy conversion from sunlight to carbohydrate, but in terms of how motion is generated that sustains the gas exchange to supply vegetation with the carbon dioxide it needs to assimilate and that simultaneously allows vegetation to evaporate water into the atmosphere. This evaporation rate from the surface to the atmosphere is strongly controlled by thermodynamics when water is sufficiently available, and this control enters twice. First, when solar radiation heats the surface, it generates buoyancy and vertical, convective motion. The more updrafts develop, the more heat and moisture is taken along from the surface into the atmosphere and carbon dioxide is replenished near the surface. With stronger updrafts, however, the surface is cooled more efficiently as well. This leads to a maximum power limit, as in the case of large-scale motion, determining the magnitude of turbulent fluxes at the surface. The second part where thermodynamics enters as a constraint is the partitioning of the absorbed radiation into heating and moistening the near-surface air. At thermodynamic equilibrium, this sets a partitioning between the sensible and latent heat flux that is known in micrometeorology as the equilibrium partitioning. The fluxes inferred from these thermodynamic constraints compare very well to observations (Kleidon et al., 2014; Conte et al., 2019). This implies that thermodynamics imposes a major constraint on the biosphere through the gas exchange of water vapour and, thus, for carbon dioxide, limiting the rate at which the terrestrial biosphere can use the absorbed solar energy in photosynthesis to generate chemical free energy.

We illustrate this reasoning with numbers from continental-scale estimates of the energy balance and the water- and carbon cycles (Stephens et al., 2012; Oki and Kanae, 2006; Beer et al., 2010) and then go into greater detail with global radiation and precipitation datasets (Loeb et al., 2018; Kato et al., 2018; Adler et al. 2016), as in Kleidon (2021b). Continental evaporation is estimated to be about $66 \times 10^{12}$ m$^3$/a (Oki and Kanae, 2006). If we assume the majority of evaporation takes place through the vegetative cover and apply the mean water use efficiency from above of 2 gC/kg H$_2$O, we obtain a gross photosynthetic uptake of $131 \times 10^{15}$ gC/a. This estimate corresponds well to the published estimate of $123 \times 10^{15}$ gC/a by Beer et al (2010). Noting that each gram of carbon in form of carbohydrate contains about 39 kJ of energy, this corresponds to a power of $152 \times 10^{12}$ W. When we then divide this power by the 165 W m$^{-2}$ of energy absorbed as solar radiation at the surface (Stephens et al., 2012) and the land area (29% of $511 \times 10^{12}$ m$^2$), we obtain a mean efficiency of the photosynthetic carbon uptake on land of 0.6%. This confirms the very low efficiency by which the biosphere generates free energy from sunlight mentioned earlier.

This estimate is, of course, very coarse, as it neglects, e.g., variations in water availability across regions. These can be seen when this analysis is done spatially more explicit using global datasets. Such an analysis is shown in Figure 5 using annual means, as in Kleidon (2021b) and as summarized in Table 1. This estimate uses the absorbed solar radiation at the surface (Figure 5a) from the CERES global radiation dataset (Loeb et al., 2018; Kato et al., 2018) as the starting point, estimates evaporation from the maximum power limit without water limitation (the so-called potential evaporation rate, Figure 5b), uses mean precipitation rate (Figure 5c) using the GPCP dataset (Adler et al. 2016) to restrict evaporation in the presence of water availability, yielding the so-called actual evaporation rate (Figure 5d).

Using the observed mean water use efficiency to the thermodynamically-derived evaporation rate then yields an estimate for the photosynthetic carbon uptake and its associated thermodynamic conversion efficiency (Figure 6). We can see that there are clear variations in efficiency among regions, with the humid regions having a greater efficiency of up to 3%, while desert regions have no productivity due to the lack of water. This supports the well-established notion that water availability is a major constraint for the terrestrial biosphere, shaping the spatiotemporal patterns of its productivity. What our estimate implies is that the limit set by gas exchange and water availability can explain very well the observed patterns of carbon uptake of the terrestrial biosphere.

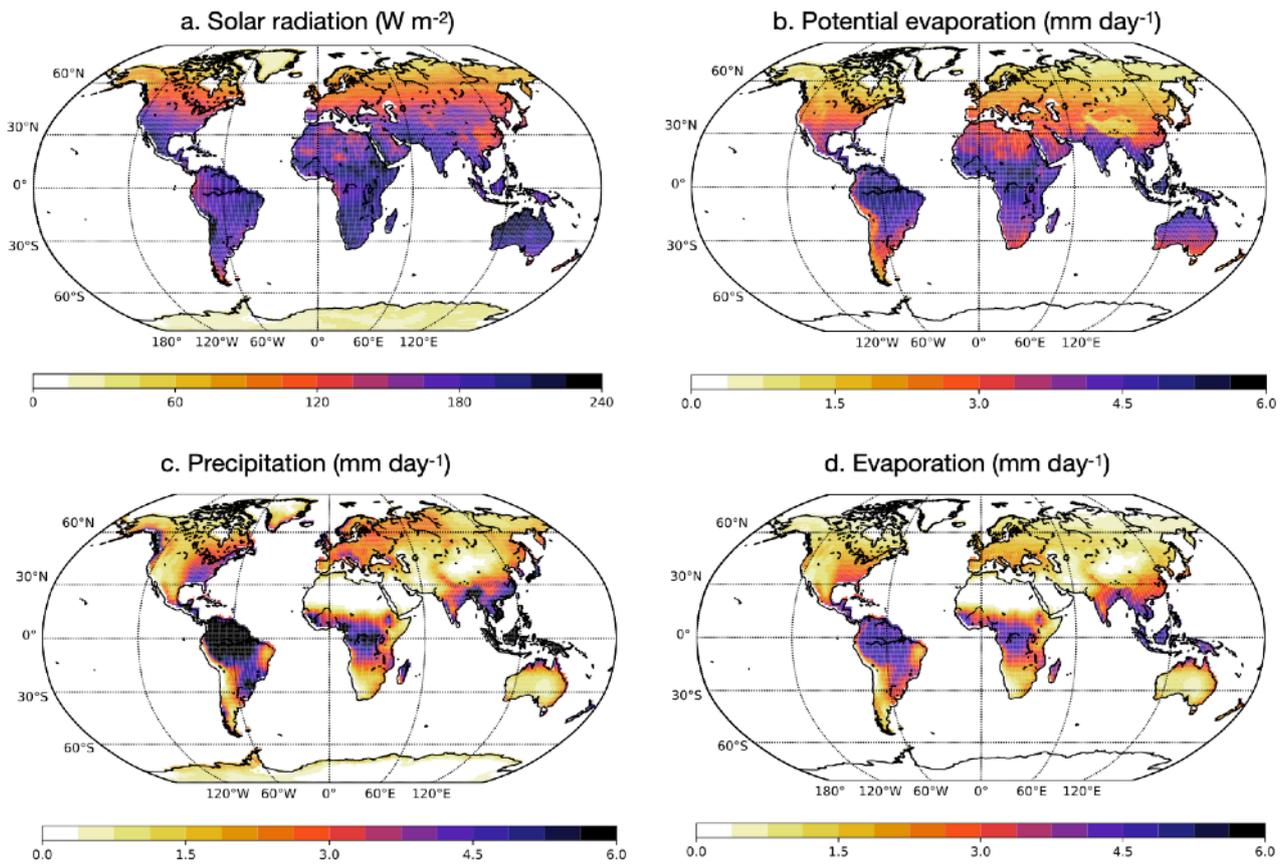

**FIGURE 5:** Estimates of mean annual evaporation rates from the thermodynamic maximum power limit and water availability. a. The solar forcing in terms of absorbed solar radiation, taken from the CERES dataset (Loeb et al., 2018; Kato et al., 2018). b. Potential evaporation rate estimated from maximum power and thermodynamic equilibrium partitioning. This rate represents evaporation when sufficient water is available. c. Mean precipitation rates, taken from the GPCP dataset (Adler et al. 2016). d. The final estimate of actual evaporation, taken as the minimum of potential evaporation and precipitation.

To conclude this part of free energy generation by the terrestrial biosphere, we note that thermodynamics does not act directly to limit the energy conversions from sunlight to carbohydrates. After all, the photosystems are highly efficient in the first steps of converting solar energy. It would rather seem that it is the rate of gas exchange that limits photosynthetic carbon uptake as it provides the necessary supply of carbon dioxide from the surrounding air. This interpretation can explain the very low efficiency in observed photosynthetic carbon uptake rates in natural ecosystems. It represents an indirect thermodynamic constraint that requires an Earth system view which describes biospheric productivity as a process that is intimately linked to, and constrained by, physical transport processes of the Earth's environment.

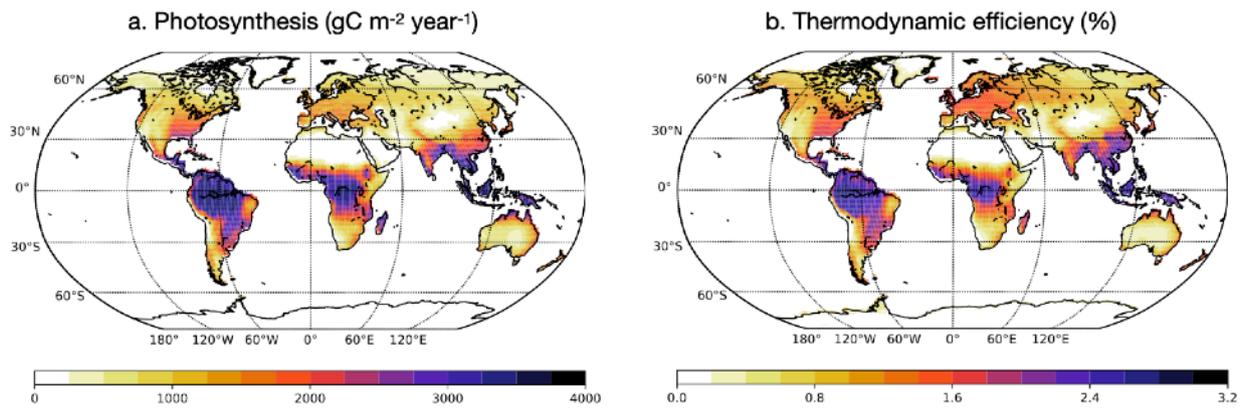

**FIGURE 6:** Physical constraints shape the productivity of the terrestrial biosphere. (a.) Gross carbon uptake by photosynthesis of terrestrial vegetation estimated from a constant water use efficiency and the thermodynamically-constrained evaporation rate shown in Figure 5. (b.) The thermodynamic efficiency in converting absorbed solar radiation into chemical free energy. Based on Kleidon (2021b).

**TABLE 1:** Annual mean energy and mass fluxes averaged over land, from the solar forcing and precipitation input to terrestrial productivity, its human appropriation, and potential means to push beyond the natural limits to these fluxes.

| | | |
|---|---:|---|
| **Natural limits of terrestrial productivity** | | |
| Absorbed solar radiation | 141 W/m² <br> 20726 TW | Calculated directly from CERES (Loeb et al., 2018; Kato et al., 2018). Shown in Figure 5a. |
| Potential evaporation | 2.84 mm/day <br> 153 x 10³ km³/year <br> 12091 TW | Calculated from the maximum power limit and thermodynamic equilibrium partitioning using CERES forcing. |
| Precipitation | 2.18 mm/day <br> 117 x 10³ km³/year | Calculated directly from GPCP (Adler et al., 2016). |
| Evaporation | 1.65 mm/day <br> 88 x 10³ km³/year <br> 7015 TW | Calculated by taking the minimum of potential evaporation and precipitation at the annual scale. Shown in Figure 5b. Compare to observed estimate of 66 x 10³ km³/year (Oki & Kanae, 2006). |
| Gross primary productivity (Net photosynthesis) | 405 gC/m²/year <br> 177 GtC/year <br> 224 TW | Calculated by converting evaporation to carbon uptake with a fixed water use efficiency of 2 gC/kg $H_2O$. Shown in Figure 5c. Compare to observed estimate of 120 GtC/year (Beer et al. 2010). |
| Net primary productivity (Biomass production) | 202 gC/m²/year <br> 89 GtC/year <br> 112 TW | 50% dissipation by autotrophic respiration by plants. |
| **Pushing the limits by seasonal soil water storage** | | |
| Evaporation without seasonal water storage | 1.47 mm/day <br> 79 x 10³ km³/year <br> 6244 TW | Calculated by assuming that monthly evaporation is the minimum of potential evaporation and precipitation (i.e., no seasonal water storage). |
| Enhancement by vegetation | 0.18 mm/day <br> +12% <br> 9 x 10³ km³/year <br> 711 TW | Calculated by assuming that evaporation is the minimum of potential evaporation and precipitation at the annual scale (i.e., seasonal water deficits are compensated by water storage variations within the rooting zone). |
| Enhancement of net primary productivity | 25 gC/m²/year <br> +12% <br> 10 GtC/year <br> 12 TW | Converted with a fixed water use efficiency of 2 gC/kg $H_2O$. |
| **Human appropriation of productivity** | | |
| Absorbed solar radiation | 111 W/m² <br> 5449 TW | Weighted average over cropland and pastures shown in Figure 7. |
| Evaporation | 1.38 mm/day <br> 24.8 x 10³ km³/year <br> 1963 TW | Weighted average over cropland and pastures shown in Figure 7. |
| Net primary productivity | 503 gC/m²/year <br> 25 GtC/year <br> 31 TW | Calculated by converting evaporation to carbon uptake with a fixed water use efficiency of 2 gC/kg $H_2O$. Reduced by 50% dissipation by autotrophic respiration by plants. |
| **Pushing limits by technology** | | |
| Runoff potentially available for irrigation and additional evaporation | 0.53 mm/day<br>29 x 10³ km³/year<br>+32% | Difference between current climatological precipitation and evaporation on land. |
| Enhancement of terrestrial net primary productivity | 193 gC/m²/year<br>28 GtC/year<br>36 TW<br>+32% | Calculated by converting evaporation to carbon uptake with a fixed water use efficiency of 2 gC/kg $H_2O$. Reduced by 50% dissipation by autotrophic respiration by plants. |
| Area needed for photovoltaics to generate current human primary energy demand of 18 TW | 550 000 km² | Calculated using the global mean absorption of solar radiation of 165 W m$^{-2}$ and a photovoltaic efficiency of 20%. |
| Area needed for generating as much freshwater by seawater desalination as is currently in continental runoff | 177 000 km² | Calculated using an energy demand of 4 kJ per litre of desalination using membrane technology and energy generation by photovoltaics using 165 W m$^{-2}$ and an efficiency of 20%. |

# 4. THE BIOSPHERE PUSHES ITS LIMITS

While the activity of the terrestrial biosphere is limited by these constraints, it nevertheless acts and affects the environment in such a way as to push these limits further to achieve higher levels of activity. The specific means and mechanisms are, obviously, different to atmospheric heat engines described above. It relates to the effects that biotic activity has on its environment, and the consequence of these for the conditions to generate and dissipate free energy. The overall dynamics of "pushing the limits" appears to reflect the same underlying evolutionary dynamics as the physical dynamics of the climate system: to maximize power and dissipation.

To illustrate this push, I want to use the depth of the rooting zone of vegetation and its effects as one specific example. As plants grow, they allocate some of their energy to grow rooting systems into the soil. A deeper rooting zone allows them to access more of the water stored in the soil for evaporation, particularly during dry periods. These periods are characterized by potential evaporation exceeding precipitation. Water stored within the soil can be used to compensate for this lack of precipitation input, allowing vegetation to maintain evaporation during such periods.

By building root systems and enhancing soil water storage, the biosphere benefits by elongating the period over which gas exchange can be maintained, and productivity be sustained. It thus makes the biosphere more productive. This enhancement is, however, not infinite, but set by the climatological water balance. In humid regions with dry periods, vegetation needs to essentially only store the water needed to overcome the water deficit during the dry season. In arid regions, vegetation cannot store more water than the water surplus during the wet season. The required water storage sizes needed for this seasonal compensation compares well to observed rooting depths in different vegetation types (Kleidon and Heimann, 1998). Figure 6 illustrates these considerations, using monthly mean fields of precipitation and the thermodynamically-constrained evaporation estimate to infer the actual evaporation rate in the absence of seasonal water storage and its enhancement by soil water storage facilitated by rooting systems. This effect of rooting systems enhances terrestrial carbon uptake by roughly 10% (Table 1), enhancing the power and dissipative activity of the terrestrial biosphere.

There are other biotic effects that can act similarly to enhance terrestrial productivity. Vegetated surfaces are typically darker (have a lower surface albedo) than bare ground, thereby enhancing surface heating as a driver for gas exchange, an aspect that has not been considered here. The highly heterogeneous canopies of forested surfaces represent a much greater leaf surface area which facilitates greater gas exchange. Stomates, small openings in the leaves to regulate gas exchange, can vary in size and numbers, and operate to maximize the carbon gain for a given water loss (Cowan & Farquhar, 1977; Medlyn et al, 2011). These means to enhance productivity all come at their environmental consequences, for instance, in terms of enhanced moisture recycling on land. On longer time scales, the activity of the biosphere has profoundly altered the atmospheric composition and the strength of the greenhouse effect, changing the planetary disequilibrium state and energy fuels for the biosphere (Judson, 2017). These effects all affect the physical environment and provide means to maximize free energy generation further, resulting in environmental conditions that sustain the current high, possibly even maximized levels of biotic activity on the planet. Even though the means by which the biosphere would achieve this maximization are very different to the physical heat engines, the outcome would be the same: to maximize free energy generation and its dissipation.

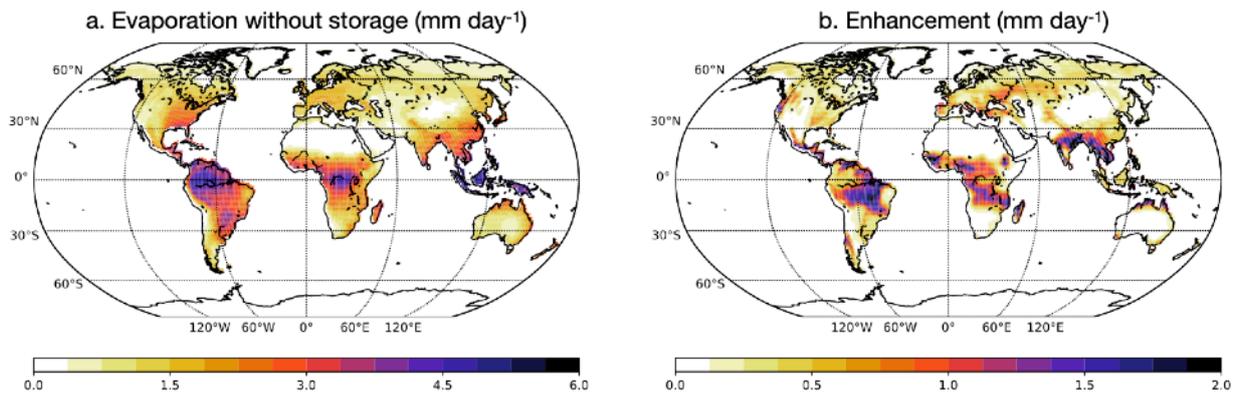

**FIGURE 7:** Effects of deep rooted vegetation on evaporation and associated gas exchange as an example of how the biosphere pushes its limits. (a.) Evaporation without seasonal soil water storage. (b.) Enhancement of evaporation due to soil water changes enabled by deep rooting systems.

## 5. HUMAN SOCIETIES AS AN ADDITIONAL ENERGY DISSIPATION TERM

I want to next turn to human activity as a thermodynamic Earth system process. To do so, we start with the consumption of free energy, which is at the very core of human existence as well as its socioeconomic activities. Humans need energy to sustain their metabolism, just like any other living organism. This energy comes in the form of the food we eat, as reflected in the calories that it contains (with calories being outdated unit of energy, with 1 cal = 4.2 J). As this energy is consumed by metabolic activity, it converts the chemical free energy associated with the disequilibrium of carbohydrates and oxygen back into carbon dioxide, water, and heat. Likewise, human societies consume free energy in form of primary energy, currently mostly in form of the chemical energy stored in fossil fuels. Upon combustion, this free energy is converted into heat, and subsequently into work, e.g., by generating motion, electricity, or transforming materials. Thus, primary energy consumption is highly correlated with economic activity (e..g, Cleveland et al., 1984; Ayres and Nair 1984). Viewing human activities primarily through the lens of energy allows us to describe it as a dissipative Earth system process and place it into the same thermodynamic framework that we utilized above for physical and biotic Earth system processes. Using this framework, we will evaluate whether human activity acts to deplete or enhance the dissipative activity of the biosphere and link this to sustaining the biosphere.

Human activity dissipates the free energy that was generated by the Earth system, specifically the chemical free energy generated by photosynthesis. Photosynthesis fuels the plants grown in agriculture, and the resulting products feed livestock and human metabolisms. A fraction of the productivity of the biosphere is thus being appropriated by agricultural activities. This share, the human appropriation of net primary productivity (Vitousek et al., 1986; Haberl et al. 2014), is considerable, estimated to be 13% to 25**%** of the total terrestrial productivity. The free energy associated with this productivity is thus diverted to direct or indirect human use, such as food production, feeding livestock, or biofuel production. It is no longer available for the natural biosphere, reducing its level of dissipative activity and sustaining less natural living organisms.

To illustrate the magnitude of human appropriation using the estimates from above, I used the land cover datasets of Ramankutty et al. (2008) as masks to describe where terrestrial productivity is appropriated by human use, either in form of croplands or rangelands (Figure 8). The estimate of carbon uptake shown in Figure 5 was reduced by 50% to account for the metabolic activity of the producers (the autotrophic respiration) and then summed up over the areas of human land use, yielding the estimates shown in Table 1. It shows that on these human-dominated areas, about 31 TW (or 28%) of the net primary productivity on land takes place, an estimate that is consistent with the more elaborate estimates by Haberl et al. (2014). Note, however, that at present, not all of this energy is appropriated to human use, as some of this energy feeds natural grazers (or "pests") or the decomposition by soil organisms, which also draw from this free energy to sustain their metabolisms. Yet, with the intensification of agricultural activity, which aims at increasing yields, this will inadvertently result in a greater share of human appropriation instead of feeding the natural biosphere. We would thus expect that with agricultural expansion and intensification, the trend of greater appropriation would continue, depleting the ability to feed the dissipative activity of the natural biosphere further.

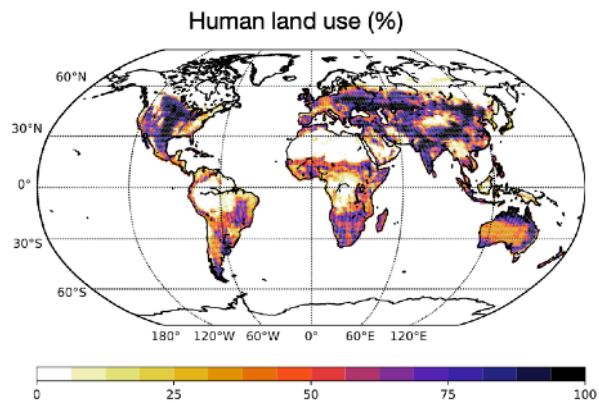

**FIGURE 7:** Human impact on the terrestrial biosphere as reflected by its land uses as croplands and rangelands. Datasets from Ramankutty et al. (2008).

The primary energy consumption due to socioeconomic activities represents further energy consumption by human societies. At present, this consumption amounts to about 18 TW, which is mostly consumed in form of fossil fuels. This chemical energy was generated by photosynthesis in the Earth's past, the subsequent burial of a fraction of the resulting biomass by geologic processes, and created the chemical disequilibrium of hydrocarbons in the geological reservoirs and atmospheric oxygen. Using fossil fuels depletes this disequilibrium, it increases the atmospheric concentration of carbon dioxide, enhances the associated greenhouse effect, and causes global warming. Irrespective of these global effects, fossil fuels are a finite resource and its use at the current rate is clearly not sustainable. If, for simplicity, we would assume that this use of energy is being replaced by appropriating more of the net primary productivity as an energy resource (e.g., in form of firewood or biofuels), it would draw away another substantial fraction of the free energy available to the natural biosphere. This would further deplete the ability to sustain the dissipative activity of the natural biosphere.

This description of human activity as an additional dissipative Earth system process would seem to suggest that more human activity would come inevitably at the expense of diminished biospheric activity. Since terrestrial productivity operates at its limit, the allocation of the generated free energy among the dissipative activity of the natural biosphere or human activity is fixed and appears to be a zero-sum game. It would then imply that the only way to improve and sustain the conditions of the natural biosphere would be for human societies to consume less energy.

## 6. SOME TECHNOLOGY PUSHES THE LIMITS TO HIGHER LEVELS

There is another option to sustain the biosphere, which relates to mechanisms to "*push the limit*", similar to the example given by which the biosphere pushes its limits. This involves certain types of human-made technologies. Examples for existing technologies are the use of river water for irrigation in arid regions or the damming of water flow to form reservoirs for irrigation during dry periods. This makes water available for enabling plant productivity at places or times in which the precipitation input is too low to meet the potential evaporation rate. The additional water made available by these technologies can act to enhance productivity by supplementing means of storing and redistributing water that were unavailable to the natural biosphere. This water can then be used to push the limit of productivity to a higher level by making more water available.

A look at Table 1 can yield us a broad estimate of the magnitude by which such technologies could, in principle, enhance terrestrial productivity by storing or redistributing water. If we take all of the continental river discharge, or runoff, which in the climatological mean balances the difference between precipitation and evaporation on land, and make it available for evaporation, by storing or redistributing it, this would enhance continental evaporation by 32%. Using the water use efficiency for conversion into a productivity increase, as was done before, this would yield an increase that is about three times as much as the enhancement of productivity due to the seasonal water storage maintained by deep-rooted vegetation. What is not accounted for here are possible climatic effects. The enhanced evaporation would clearly result in more continental moisture recycling, cloud cover, and precipitation, and thus change the environmental conditions on land. Nevertheless, this example is made simply to show that already existing technology can provide alternative means to enhance productivity and its human appropriation that does not come at the cost of appropriating more of the natural productivity of the biosphere.

When we look into the future, a far bigger effect can be achieved with modern technology. Photovoltaics provides a technology that generates free energy directly from sunlight at much greater efficiency than heat engines or photosynthesis can ever achieve. By converting solar radiation directly into electricity, it avoids the inevitable, irreversible losses by the conversions into heat, as is the case for the heat engines of the atmosphere, and it is not constrained by gas exchange and water availability, as is the case for photosynthesis, because photovoltaics exports its free energy in form of electricity, not needing gas exchange. With photovoltaics, human societies can thus become a producer of free energy of the Earth system, and thereby decouple their demand from the supply by the biosphere. In other words, human societies can sustainably grow further for some time, but this does not need to come at the expense of the biosphere.

It would require relatively little area to meet the current demands for primary energy by photovoltaics: With a typical efficiency of about 20% for solar panels and a mean absorption of solar radiation of 165 W m$^{-2}$, it would merely require about 550 000 km$^2$ or less than 0.4% of the land surface to meet the current primary energy consumption. The use of photovoltaics would thus take away the pressure imposed by meeting the primary energy consumption from the appropriation of energy from the biosphere, as fossil fuels or biofuels, or from the renewable energy generated directly or indirectly by the heat engines of the atmosphere, such as wind or hydropower.

This novel supply of primary energy can then be supplemented by other technologies to alleviate other natural limits of the biosphere, particularly the ones imposed by water availability. Seawater desalination using membrane technologies requires a very small fraction of the energy involved in the natural desalination process by evaporation and subsequent precipitation. While it takes about 2.5 MJ to evaporate and desalinate a litre of seawater (known as the latent heat of vaporisation), membranes only require about 4 kJ to achieve the same result (Elimelech and Phillip, 2011). To put these numbers in a global perspective: Currently, it requires 3650 TW of solar energy, or 3% of the absorbed solar radiation to evaporate water to feed the net convergence of moisture transport to land of 46 x 10$^3$ km$^3$ per year (Oki & Kanae, 2006; estimate in Table 1 is 29 x 10$^3$ km$^3$ per year). To obtain the same rate of freshwater production by seawater desalination using membrane technology, it would require 6 TW of energy, which could be achieved by photovoltaics installed over 177 000 km$^2$ of area (using global means).

These are, of course, rough estimates that do not take into account the many practical challenges to make this happen. Changing the terrestrial hydrologic cycle at this order of magnitude would clearly result in climatological changes, likely enhancing continental precipitation. Yet, the point I want to make with these estimates is that there are a few human-made technologies already available that achieve the outcome of natural processes with much greater efficiency. This, in turn, could decouple the growth in food and energy needs of human societies from their natural sources, decreasing the magnitude in appropriation while potentially resulting in positive feedbacks on photosynthetic carbon fixation (cf. Figure 1). This decoupling could reduce the impact on the natural biosphere by allowing it to use its free energy to feed the dissipative activity of its natural food webs, and thus sustain the activity of the natural biosphere at higher levels.

## 7. A SUSTAINABLE FUTURE OF THE TERRESTRIAL BIOSPHERE

I used a thermodynamic Earth system perspective to evaluate how the activity of the natural biosphere can be sustained in the presence of increasing human activities. I first reviewed the application of thermodynamics to show how much it constrains the physical functioning of the climate system, and thereby the activity of the terrestrial biosphere. This results in a basic trade-off: increased human appropriation of energy seems to come inevitably at the cost of reducing the dissipative activity of the natural biosphere. The way out of this dilemma is the use of novel technology, particularly photovoltaics. This allows human societies to generate free energy from sunlight more efficiently than natural means, particularly on areas that are currently not generating free energy, such as deserts. The use of this energy can then decouple human energy needs from the supply by the natural biosphere. It is through this decoupling that human activity could, in principle, grow sustainably further to some extent, with this growth coming not at the expense of shrinking the natural biosphere further, but providing a possibility to sustain and even enlarge the natural biosphere in the Anthropocene.

Such a trajectory of sustained further growth would likely lead to quite a different physical environment. When this energy is used to generate more resources such as freshwater to extend agriculture into arid regions, instead of further deforesting humid regions, it would simultaneously strengthen hydrologic cycling and thereby alter the physical climate system. Yet, human activities consume energy at rates of similar magnitude to natural processes. It is hard to imagine that this consumption would voluntarily be drastically reduced in the future. With this constraint, it would seem inevitable that to preserve the natural biosphere, the only option that human societies would have is to "enlarge" the biosphere into areas that are currently not productive, such as desert regions, in order to sustain the dissipative activity of the natural biosphere at current levels.

I hope that this energy-focused view of the biosphere and sustainability of human activity at the very large, planetary scale can be useful as an inspiration for practical applications to evaluate human interactions and how detrimental or beneficial these may be for the natural biosphere to persist in times of greater human influences.

**Data availability:**
The datasets used to create the figures and to make the estimates shown in Table 1 will be made available upon acceptance of this manuscript.